\documentclass[12pt,preprint]{aastex}

\newcounter{sub}
\newcounter{subeqn}[sub]
\newcommand\fzA{f_0^{A}}
\newcommand\fpA{f_+^{A}}
\newcommand\fmA{f_-^{A}}

\newcommand\zetalamb{\zeta^{A{\Lambda}}}	
\newcommand\chialamb{\chi^{A{\Lambda}}}	
\newcommand\gammaalamb{\gamma^{A{\Lambda}}}

\newcommand\ofr{(r)}
\newcommand\unit{{\mbox{\boldmath $e$}}}
\newcommand\opL{{\bf L}}

\newcommand\bzlamb{b_0^{\Lambda} }
\newcommand\bplamb{b_+^{\Lambda} }
\newcommand\bmlamb{b_-^{\Lambda} }

\newcommand\grad{{\mbox{\boldmath $\nabla$}}}

\newcommand\be{\begin{equation}}
\newcommand\ee{\end{equation}}

\newcommand\lp{\left(}
\newcommand\rp{\right)}

\newcommand\ls{\left[}
\newcommand\rs{\right]}

\newcommand\st{\stepcounter{sub}}
\newcommand\stq{\stepcounter{subeqn}}
\newcommand\bea{\begin{eqnarray}}
\newcommand\eea{\end{eqnarray}}

\newcommand\xxi{{\mbox{\boldmath $\xi$}}}

\newcommand\nab{\mbox{\boldmath $\nabla$}}
\newcommand\cchi{{\mbox{\boldmath $\chi$}}}
\newcommand\ggamma{{\mbox{\boldmath $\gamma$}}}

\newcommand\B{{\bf B}}
\newcommand\E{{\bf E}}
\newcommand\J{{\bf J}}
\newcommand\C{{\cal C}}
\newcommand\W{{\cal W}}
\newcommand\F{{\bf F}}

\newcommand\no{\nonumber}

\begin{document}

\title{Normal Modes of Rotating Magnetic Stars}
\author{S. M. Morsink$^1$ \& V. Rezania$^{1,2}$}
\affil{$^1$Theoretical Physics Institute,
	Department of Physics,
	University of Alberta\\
	Edmonton, AB, Canada, T6G 2J1}

\affil{$^2$Institute for Advanced Studies in Basic Sciences, 
          Zanjan 45195, Iran}

\begin{abstract}

We investigate the effect of a magnetic field on the
global oscillation modes of a rotating fluid star in the
magnetohydrodynamic approximation. We present general equations
for the modification of any type of fluid mode due to a general
magnetic field which is not aligned with the star's spin axis. 
In the case of any internal dipole magnetic field we 
derive the equations for the frequency corrections to the
r-modes. We solve for the frequency correction explicitly for
the case when the internal dipole field is force-free,
including the uniform density case.
In the weak-field limit,
the spatial form of the r-mode velocity perturbation
is unchanged, but the magnitude of the frequency
in the rotating frame increases.

\end{abstract}


\section{Introduction}

The observation of oscillations of the Sun and other
stars has led to great advances in our understanding of the
internal structure of stars. While the oscillations of 
non-rotating, non-magnetic stars are well understood, 
less is known about the oscillations of stars which 
are rotating and possess a magnetic field \citep{Cox80,Unn89}. 
The main classes of stars for which magnetic fields and rotation
both play important roles are the rapidly oscillating Ap (RoAp) stars,
the magnetic white dwarf stars and neutron stars. The Ap stars 
are hydrogen burning stars with strong magnetic fields. Rapid 
oscillations (relative to the spin frequency)  have been observed 
in the subclass of stars known as RoAp stars \citep{Kur90}.
These oscillations have been identified as high overtone 
p-modes. In addition, evidence for line splitting by rotation
has been found. These stars are modeled with a magnetic field
tilted with respect to the spin axis of the star. 

Numerous white dwarf stars have been observed to oscillate
\citep{GS96}. The oscillations of the white dwarfs
have been identified as g-modes. So far no magnetic field larger
than $\sim kG$ \citep{SG97} has been observed
for any of the pulsating white dwarfs. On the other hand, a
class of white dwarfs with magnetic fields up to $10^9 G$
\citep{WF00}
known as magnetic white dwarfs (MWD) has shown no evidence 
of pulsations. The lack of pulsations in the MWDs may be due
to magnetic suppression of pulsations or to pulsation 
amplitudes which are not large enough to be detectable.

The observation of quasiperiodic pulsar microstructure 
\citep{Bor76,Cor90} has led to the suggestion \citep{VH80}
that neutron star oscillations had been observed.
While the evidence for electromagnetic detection of
neutron star oscillation modes is inconclusive, it
may be possible that neutron star oscillations will be detectable by
advanced gravitational wave detectors. 
It has also been proposed that the soft 
gamma repeaters (SGR) are high magnetic field neutron 
stars with fields as high as $10^{15} G$ \citep{TD95}.
This hypothesis is supported by observations of objects
such as SGR 1900+14 which appears to have a magnetic 
dipole field strength in the range $\sim 2-8\times 10^{14}G$
\citep{Kou1998}. Torsion modes in the crusts of these
neutron stars may have been detected in SGR outbursts
\citep{Dun98}.

The relative importance
of magnetic fields and rotation in different types of 
stellar systems varies by many orders of magnitude, as can be 
seen in Table 1, where the ratio of magnetic field energy,
${\cal{M}}= B_{in}^2 R^3/6$ to rotational kinetic energy, 
$T=MR^2(2\pi/P)^2/5$, is given for typical stars. 
In Table 1 we give values for the average magnetic field,
and the magnetic field energy is calculated as though the star's 
magnetic field is uniform. For
 millisecond pulsars and typical white dwarf stars,
rotation is much more important than magnetic fields at global
scales. At the other end of the scale, if the soft gamma repeaters
can be modeled as high magnetic field neutron stars, magnetic 
fields are dominant over rotation. The magnetic white dwarf stars
and some high magnetic field neutron stars have both magnetic 
field and rotation energies of similar sizes. Clearly any
mathematical method which treats the ratio of magnetic field energy
to rotational energy as a small perturbation can not 
describe the phenomenology of all types of stars listed in 
Table 1.

In this paper we introduce a formalism for computing global
oscillation modes of magnetic, rotating Newtonian fluid stars. 
Our method makes use of a slow-rotation expansion, but 
all other aspects are nonperturbative, in that the modes
of stars with any value of the ratio of ${\cal{M}}/T$ can be 
computed. The method of computation is particularly useful
for the computation of magnetically modified r-modes
and inertial modes which vanish in the limit of zero rotation. 
For illustrative purposes the magnetic corrections
to the r-modes of a constant density
star with a homogeneous magnetic field inclined at arbitrary
angle to the star's spin axis will be presented. In the 
example presented, the computed modes are only valid in 
the limit of small ${\cal{M}}/T$.  The full treatment of
the modes of stars with strong magnetic fields is more
involved than our example and will be presented elsewhere. 
It is straight-forward to extend these calculations to stars
with realistic equations of state and more complicated
magnetic fields. 

There is already an extensive literature on the pulsations
of magnetic stars. Radial oscillations of nonrotating magnetic stars
were first investigated by \citet{CF53}. \citet{LS57} computed the
magnetic corrections to the nonradial f-modes of a nonrotating star.
In these early calculations, the overall effect of the magnetic 
field was assumed to be weak, in the sense that the energy stored
in the magnetic field is small compared to the star's gravitational
binding energy. While this is generally true for all stars of interest,
the magnetic pressure can dominate over gas pressure near the star's surface.
This suggests that a perturbative approach will break down near the 
surface of a star. \citet{Car86} introduced a cylindrical model of the
region near the magnetic pole at the surface of a neutron star and computed 
the spectrum of nonradial oscillations of the model problem. A magnetic
boundary layer approach was adopted by \citet{DG96} and \citet{Big00} 
as an alternative method for finding the magnetically modified modes 
near the stellar surface.

Computations of nonradial oscillations of rotating magnetic stars 
are more complicated than the calculation of modes of magnetic 
nonrotating stars. One approach is perturbative: the star's rotation
and magnetic field can be taken to be small corrections to the
modes of nonrotating, nonmagnetic stars. This perturbation approach
has been applied to the calculation of magnetically modified 
pressure modes of rotating stars \citep{ST93,TS94,TS95}. 
The work on modes of rotating magnetic stars has so far focused
on the pressure modes. However, in the absence of a magnetic field
a rotating star has another family of modes, the inertial modes,
which are driven by the Coriolis force. Malkus (1967) computed
the magnetic generalization of these modes for the case of
a star whose magnetic field is generated by an electric current
which is parallel to the fluid's angular velocity. In this 
special case the equations describing normal modes can be 
separated in oblate spheroidal coordinates. It is of interest,
however, to understand the magnetic generalizations of the
inertial modes for magnetic fields which are tilted 
with respect to the star's rotation axis, as is the
situation common in astrophysics. If excited in the Ap stars, these modes 
may create beat frequencies with the high order p-modes which 
may be observable. (The inertial modes have a very small radial 
component to their oscillation and are unlikely to provide
an observable luminosity variation on their own.) Another 
motivation for computing the magnetically modified 
inertial modes is due to the prediction that it may be possible
for a subset of the inertial modes, the r-modes, to be
driven unstable by gravitational radiation \citep{And98,FM98,LOM98}
in young neutron stars. In a series of papers \citet{Rez00,Rez01a,Rez01b}
have examined the consequences of a large amplitude r-mode and shown
that it would have the effect of amplifying the star's magnetic field
and hindering the further growth of the mode. 
This effect could play an important role in the early evolution 
of a neutron star's spin and magnetic field. Further work by
\citet{HL00} provided further insight on the effect of a growing r-mode on 
the magnetic field and calculated the form of magnetically modified
r-modes in a plane-wave approximation. One of the motivations for the present
paper is to provide a framework for computing the form of the magnetically 
modified global r-modes. 
The formalism and calculations presented in this paper are only for
global oscillation modes of an infinitely conducting fluid. We have not 
included features such as a crust or a magnetosphere which would 
exist in the case of neutron stars. It should be possible to extend
the present formalism to include interactions with crustal oscillation
modes and a magnetosphere. 

The structure of this paper is as follows. In section \ref{mhd} we review the
equations describing the modes of rotating magnetic fluid stars.
 The form of
the equilibrium magnetic field is discussed in section \ref{equilmag}.
In section \ref{rmodes} we show how the frequencies of the r-modes
are modified by the presence of a dipole magnetic field.
We examine two model interior fields, the uniform magnetic field
and the force-free magnetic field.
In the case of a uniform magnetic field
the frequency in the rotating frame of an r-mode 
with quantum number $\ell$ is
\st\be
\sigma = -\frac{2\Omega}{\ell+1}
\left( 1 - \frac{\cal{M}}{T} 
\frac{1}{10} (\ell+1)(2\ell+3) \left[ 1
+ \frac12 (\ell(\ell+1)-3) \sin^2\alpha\right]\right),
\ee
where $\Omega$ is the angular velocity of the star,
$\cal{M}$ is the energy stored in the magnetic field,
$T$ is the star's rotational kinetic energy and
$\alpha$ is the angle between the magnetic field's
axis and the spin axis.  
In the case of the force-free magnetic field, numerical results are
presented.
Finally, we conclude in section \ref{conclusions} with remarks about 
further problems which can be attacked with the present formalism.

\section{Normal Modes of Rotating Stars with Magnetic Fields}\label{mhd}

In this paper we study the magneto-hydrodynamic (MHD) perturbations
of a uniformly rotating star endowed with a magnetic field. 
As we do not assume that
the symmetry axis of the magnetic field is aligned with the star's 
rotation axis, magnetic multipole radiation is emitted, and 
no equilibrium solution exists. However, we only consider
timescales short compared to the characteristic stellar spin-down time 
due to the nonaxisymmetric magnetic field. In this sense an equilibrium
solution exists. We work within the ideal MHD framework \citep{Jac75}
where the star's fluid is assumed to be electrically neutral and
to have infinite conductivity. With the assumption of infinite conductivity,
the magnetic field lines are frozen into the fluid, and the magnetic
field pattern rotates at the same rate as the star. 

The equilibrium star is described by the density $\rho$, fluid
pressure $p$, gravitational potential $\phi$, velocity
$\bf{v}$ and magnetic field $\B$. Eulerian perturbations of these
quantities are denoted with the symbol $\delta$.
For infinite conductivity, the Eulerian perturbations of
the magnetic field and current density are
\st
\begin{mathletters}
\bea
\label{delta-B}
{\partial\over\partial t}\delta\B &=&\nab\times(\delta{\bf v}\times\B)\\
\label{delta-E}
\delta\E&=&-{1\over c}\delta{\bf v}\times\B\\
\label{delta-J}
{1\over c}\delta\J&=&{1\over 4\pi}\lp\nab\times\delta\B-{1\over c}
{\partial\over\partial t}\delta\E     \rp,
\eea
\end{mathletters}
where $\delta {\bf v} = \partial_t \xxi$ and the perturbed 
density, pressure and gravitational field are given in
standard works on perturbation theory such as \citet{Unn89}.
The fluid displacement
vector $\xxi$ is a solution of Euler's equation, 
\st\be\label{euler-1}
\ddot\xxi+2\C\dot\xxi+\W\xxi=\frac{1}{\rho}\F^{mag}\,,
\ee
where the operators $\C$ and $\W$ are defined by
\st
\be
\label{C-W}
\C\xxi\equiv \Omega\times\xxi\,, \quad
\W\xxi\equiv
\frac{1}{\rho} \nab\delta p -
{\nab p \over \rho^2 } \delta\rho + \nab \delta\phi \,,
\ee
and  the magnetic force is
\st\be\label{F-mag}
\F^{mag}= \frac{1}{c}\left(
\delta\J\times\B + \J\times\delta\B\right).
\ee
The possible perturbations of a magnetized rotating star are
found by specifying the equilibrium configuration of the
star and then solving equation (\ref{euler-1}). This is 
a nontrivial problem, and the rest of this paper will be 
concerned with a method for finding normal mode solutions
of Euler's equation.

In this paper we adopt the method introduced by 
\citet{Sch01} where the perturbations are expanded in the
spatial eigenfunctions of Euler's equation with zero
magnetic field.
The normal modes of rotating stars with no magnetic 
fields (or any other external forces) are solutions of Euler's equation
(\ref{euler-1}) with $F^{mag}=0$ of the form 
$e^{-i\omega_At}{\bf \xi}_A(x)$ where the spatial eigenfunction 
${\bf \xi}_A(x)$ satisfies the equation
\st\be\label{euler-2}
-\omega_A^2  \xxi_A  -2 i \omega_A \C \xxi_A + \W\xxi_A=0\,.
\ee
Upper-case Latin subscripts are used to label solutions and 
correspond to the unique set of quantum numbers which describe 
the solution. We will make use of the inner product defined by
\st\be
< \xxi_A, {\cal O} \xxi_B > = \int d^3 x \rho(x) 
	\xxi_A^* \cdot {\cal O} \xxi_B ,
\ee
for any operator $\cal{O}$. 
An important property of the modes of rotating stars
discussed by \citet{Sch01} is that the modes are not
orthogonal with respect to the usual inner product. 
Instead, the modes obey the relation
\st\be
2 \epsilon_A \delta_{AB} = 
	\omega_A (\omega_A+\omega_B)< \xxi_B,\xxi_A>
	+ 2 \omega_A <\xxi_B,i\C \xxi_A> 
= \omega_A \omega_B < \xxi_B,\xxi_A> + \omega_A <\xxi_B,\W \xxi_A>
\ee
which defines 
$\epsilon_A$, the rotating frame energy of the mode
\citep{FS78}.

The lack of a simple orthogonality relation for the modes of
rotating stars has the following consequence. If a general perturbation
is expanded into a sum over modes of the form
$
\xxi(x,t) = \sum_B a_B(t) \xxi_B(x),
$
the expansion coefficients are not given by the usual formula
$a_A(t) \propto <\xxi_A, \xxi(x,t)>$ as would be true in the 
case of zero rotation. An alternative phase space expansion
has been introduced by \citet{Sch01} to circumvent this 
complication. In the phase space approach, the perturbation
and its first time derivative have expansions of the form
\st
\be
\xxi(x,t) = \sum_B q_B(t) \xxi_B(x),\label{modesum1}
\quad \hbox{and} \quad
\dot{\xxi}(x,t) = -i \sum_B \omega_B q_B(t) \xxi_B(x),
\ee
where the expansion coefficients are given by \citep{Sch01}
\st
\be
\label{coeff}
q_{_A}(t)= \frac{\omega_A}{2\epsilon_A}
\left<\xxi_A,\omega_A\xxi({\bf r},t)+i\dot\xxi({\bf r},t)
+2i\C\xxi({\bf r},t)\right> .
\ee

The equations of motion for the
expansion coefficients take on the simple form derived 
by \citet{Sch01}
\st\be\label{coeff-eq}
\dot{q}_A(t)+i\omega_A q_{_A}(t)={i\omega_A\over
2 \epsilon_A} \left<\xxi_A, {1\over\rho}\F^{mag}[\xxi({\bf r},t)]\right>
={i\omega_A\over
2 \epsilon_A} \sum_B q_B  \left<\xxi_A, {1\over\rho}\F_B[\xxi_B({\bf r})]\right>
\,,
\ee
where the magnetic force (\ref{F-mag}) has an expansion of the form
$F^{mag} = \sum_A q_A(t) F_A[\xxi_A({\bf r})]$.

The term $\F_A$ is the Lorentz force created by the fluid motion $\xxi_A$.
We now define dimensionless magnetic coupling coefficients through
\st\be
\label{def-kappa}
\kappa_{AB} = \frac{1}{\cal M} <\xxi_A, \frac{1}{\rho}\F_B>.
\ee
The integral $<\xxi_A, \frac{1}{\rho}\F_B>$ is the work done
by the force $\F_B$ when the fluid is displaced through the 
distance $\xxi_A$. The coefficients $\kappa_{AB}$ can then
be thought of as the ratio of work done by the perturbed Lorentz
force to the total magnetic energy stored in the equilibrium star.
The diagonal entries of the matrix $\kappa$ correspond to 
the work done against the fluid motion $\xxi_A$ which generates 
the Lorentz force $\F_A$. Since the magnetic forces will tend
to oppose the motion of the fluid, we expect that the diagonal
entries in the matrix of coupling coefficients will be 
negative.

After integrating by parts, the coupling
coefficients take on the form
\st\bea\label{KAD}
&&\kappa_{AD}=
- \frac{1}{4\pi{\cal M}} \int \left( \kappa^{(1)}_{AD}(r)
- \kappa^{(2)}_{AD}(r) - \frac{\omega_D^2}{c^2} \kappa^{(3)}_{AD}(r)
\right) r^2 dr\\
\stq\label{KAD-1}
&&\kappa^{(1)}_{AD}(r)=\int \nab\times(\xxi^*_A\times\B)\cdot
\nab\times(\xxi_D\times\B)~
d\Omega,\\
\stq\label{KAD-2}
&&\kappa^{(2)}_{AD}(r)=\int (\xxi^*_A\times (\nab \times \B))\cdot
\nab\times(\xxi_D\times\B)~ d\Omega
\\
&&\kappa^{(3)}_{AD}(r)=
\int (\xxi^*_A\times\B)\cdot(\xxi_D\times\B)~ d\Omega.
\stq\label{KAD-3}
\eea
The diagonal elements $\kappa^{(1)}_{AA}$ and $\kappa^{(3)}_{AA}$
are positive definite. 
In most cases $\omega R \ll c$ so that the term proportional 
to $\kappa^{(3)}_{AD}$ can
be neglected. This last term is large typically only for neutron
stars. However, since we are using Newtonian gravity, the errors
incurred in neglecting $\kappa^{(3)}_{AD}$ are similar to the
errors coming from the neglect of general relativity. Although
most authors neglect the term $\kappa^{(3)}_{AD}$, we have chosen
to keep it explicitly in our equations for the sake of completeness.

Normal mode solutions to the MHD perturbation equations can be found
by making the ansatz
\st\be
q_A(t) = c_A \exp(-i\sigma t)
\label{def-c}
\ee
where the $c_A$ are constants and $\sigma$ is the mode frequency
in the rotating frame. Substituting the ansatz (\ref{def-c}) 
into the equations of motion, we reduce the equations to the form
\st\be\label{coeff-mag-1}
\omega_A \left( c_{A}- \frac{\cal M}{2\epsilon_A}
\sum_D \kappa_{AD} c_D \right) =\sigma c_{A},
\ee
which is an eigenvalue problem for the frequency $\sigma$
and eigenvector components $c_A$.  
In order to solve the eigenvalue problem, we must truncate 
the system to a finite number of modes. 
Suppose, for instance, that we chose to truncate the 
basis to a set of N eigenmodes of the non-magnetic system.
Solving the system of equations (\ref{coeff-mag-1}) will
result in N eigenvectors and eigenfrequencies of the 
magnetic system. The dimension of the vector space can then
be increased by one by adding one more non-magnetic basis
vector and solving for the N+1 magnetic modes. The difference
between the original N frequencies can then be calculated
and the process iterated until the frequency differences
between successive iterations is suitably small. 
This procedure would be difficult to implement
 if it happened that all of the modes were coupled to each other. 
However, in the case of dipole magnetic fields, 
it will be shown in section 4 that the angular momentum selection
rules restrict the possible mode couplings so that 
each mode couples to only a small number of other modes.
As a result,  most of the $\kappa_{AD}$ coupling coefficients
vanish. This result will continue to hold for magnetic fields
of higher mulitpoles.

The constants $\epsilon_A$ appearing in the eigenvalue
equation (\ref{coeff-mag-1}) are the rotating frame energies of
the modes of a nonmagnetic star. These energies depend on
the normalization chosen for the modes. Since the problem 
of finding modes of a magnetized star is linear, the final
solution corresponding to an eigenfrequency $\sigma$ 
and eigenvector components $c_A$ is independent of the
normalizations chosen for the spatial eigenfunctions of
the nonmagnetized star. This allows us to choose any 
convenient normalization scheme which simplifies the 
form of the eigenvalue equation. A natural choice of normalization
results by noting that the rotating frame energy is
proportional to the square of the nonmagnetized mode
frequency $\omega_A$. The mode frequencies of rotating nonmagetized
stars fall into two classes, frequencies which 
in the nonrotating limit are finite (such as f-modes,
p-modes and g-modes) and frequencies which vanish
in the nonrotating limit (inertial modes, such as the
r-modes). A simple scheme is to choose the
normalization of all modes with finite frequency in the
nonrotating limit so that their energy is 
$\epsilon_A = |W|$ where $W$ is the gravitational potential
energy of a constant density star with the same mass and radius
as the star under study. Similarly, a suitable normalization is
to choose the energy of the inertial modes to be
$\epsilon_A = T$ where $T$ is the rotational kinetic energy
of the equivalent constant density star. This choice of 
normalization suggests the following approximation scheme
for weak magnetic fields. When ${\cal{M}}/T\ll1$ and 
${\cal{M}}/|W|\ll1$  the off-diagonal entries in the matrix
\st\be
N_{AD} = \delta_{AD} - \frac{\cal{M}}{2\epsilon_A} \kappa_{AD}
\ee
will be much smaller than the diagonal entries. When
$N_{AD}$ can be approximated as diagonal, 
$N_{AD} \simeq \delta_{AD} \left( 1 -  \frac{\cal{M}}{2\epsilon_A} \kappa_{AA}
\right)$
the eigenvalue problem is greatly simplified, yielding spatial
eigenfunctions identical to the eigenfunctions of the 
nonmagnetic star. In the case of modes which vanish in the non-rotating limit,
such as the r-modes, only one mode exists for each spatial 
eigenfunction. In this case, when matrix $N_{AD}$ can
be approximated by a diagonal matrix, the
equation for the MHD frequency is simply given by
\st
\be
\sigma = \omega_A \left( 1 - \frac{ {\cal{M}} \kappa_{AA}}
{2 T} \right).
\ee
If the diagonal terms in the coupling matrix $\kappa$ 
are purely real, as
in the case of a force-free magnetic field, the resulting
frequency $\sigma$ will also be real. In the case of 
r-modes of stars with force-free magnetic fields,
 the term $\kappa^{(2)}_{AA}(r)$
vanishes and the diagonal entries $\kappa_{AA}$ are all
negative definite. As a result, the magnetic field 
increases the absolute value of the r-mode frequencies. 

Implicit in this method is the assumption that the spatial
eigenfunctions of a rotating star form a complete basis.
However, it has been shown by \citet{DS79} that the normal mode of a rotating
star are complete only in a weak sense. \citet{DS79} have shown that
when the frequencies of the normal modes are all real, there are
no exponentially growing solutions of the initial-value problem. In the
case of purely real frequencies it is then possible to express
perturbations in terms of normal modes. \cite{DS79} have stressed
that their result allows the approximation of an infinite-dimensional
problem by a finite-dimensional problem. These results on completenesss
are only valid when the eigenfunctions are square integrable. 
In some special cases, such as an incompressible perfect fluid
confined to a spherical shell, it has been shown \citep{RGV01}
that there exists perturbations which are not square integrable
and that only the r-modes are regular solutions. For the case
of a spherical shell, an expansion in normal modes will not
represent all perturbations.

Clearly, this method relies heavily on the computation of
the magnetic coupling coefficients $\kappa_{AD}$.
This is not a simple task.
In the following
sections we discuss selection rules and strategies for
computing the coupling coefficients.

\subsection{Spin-weighted Mode Decomposition}

The calculation of the $\kappa_{AD}$ coupling coefficients is 
nontrivial and requires a systematic strategy. In our computation
we make use of spin-weighted spherical harmonics
and the formalism introduced by \citet{Sch01}. 

To begin with, we introduce a complex basis 
$\{ \unit_0,\unit_+,\unit_-\}$ related to the usual orthonormal
basis $\{ \unit_{\hat r},\unit_{\hat \theta},\unit_{\hat \phi}\}$ by
\st
\be
\unit_{0} = \unit_{\hat r}, \quad \unit_{\pm} 
= \frac{1}{\sqrt{2}} \left( \unit_{\hat \theta} \pm i \unit_{\hat \phi}\right).
\ee
These unit vectors have the following scalar products
\st
\be
\unit_0 \cdot \unit_0 = \unit_+ \cdot \unit_- = 1, \quad
\unit_0 \cdot \unit_\pm = \unit_+\cdot\unit_+ = \unit_-\cdot\unit_- = 0.
\ee
Any vector $\xxi$ has an expansion of the form
\st
\begin{mathletters}
\bea
\xxi = \xi_0 \unit_0 + \xi_- \unit_+ + \xi_+ \unit_- && \label{xispin}\\
\xi_0 = \xi_r, \quad \xi_\pm = \xxi \cdot \unit_{\pm} = 
\frac{1}{\sqrt{2}} \left(\xi_\theta \pm i \xi_\phi \right),
\eea
\end{mathletters}
where $\xi_r, \xi_\theta, \xi_\phi$ are the components of the vector in
the usual orthonormal basis.

The spherical harmonics with spin-weight $s$ are defined by \citep{Cam71}
\st\be
_s\!Y_{\ell m}(\theta,\phi) = \sqrt{\frac{(2\ell+1)}{4\pi}} 
	d^\ell_{-sm}(\theta) e^{im\phi}
\ee
where the $d^\ell_{sm}(\theta)$ are the matrix representations for
rotations through an angle $\theta$ discussed in 
detail by \citet{Edm74}. When the spin-weight $s=0$, the spin-weighted
spherical harmonics reduce to the regular spherical harmonics.
From the definitions of the spin spherical harmonics, it
follows that \citep{Sch01}
\st
\begin{mathletters}
\bea
\unit_{\pm} \cdot \nab Y_{\ell m} 
	&=& \mp \frac{1}{\sqrt{2}r} \sqrt{ \ell (\ell+1)} \;_{\pm1}\!Y_{\ell m} 
	\label{grad}\\
\unit_{\pm} \cdot \opL Y_{\ell m} 
	&=& - \frac{1}{\sqrt{2}} \sqrt{ \ell (\ell+1)} \;_{\pm1}\!Y_{\ell m} 
	\label{curl},
\eea
\end{mathletters}
where $\opL=-i{\bf r}\times\grad$.

The spatial eigenmodes of a rotating, unmagnetized star have azimuthal
angular dependence $\exp(im_A\phi)$, but in general do not have
a definite quantum number $\ell$. The modes can be written in
the general form (see, for example \citet{LF99})
\st\be
        \xxi_A(x) = \sum_{\ell=\vert m_A\vert}^\infty
        \left[ {W_{\ell m_A}\ofr\over r}Y_{\ell m_A}
     \unit_{\hat r} +
                {V_{\ell m_A}\ofr}
        \grad Y_{\ell m_A}-
                {U_{\ell m_A}\ofr\over r}\opL Y_{\ell m_A}\right].
    \label{genformmodes}
\ee
Making use of equations (\ref{grad}) and (\ref{curl}), the general mode expansion
(\ref{genformmodes}) is
\st
\be
\xxi_A(x) =  \sum_{\ell'} \left(
	f_{0}^{\ell m_A}(r) _0\!Y_{\ell m_A}  \unit_0  +
	f_{+}^{\ell m_A}(r) _{+1}\!Y_{\ell m_A} \unit_- +
	f_{-}^{\ell m_A}(r) _{-1}\!Y_{\ell m_A} \unit_+ \right)
\label{xif}
\ee
where the functions $f^{\ell m_A}_s$ are given by 
\st
\be
f_{0}^{\ell m_A}(r) = \frac{W_{\ell m_A}}{r} \quad \hbox{and} \quad
f_{\pm}^{\ell m_A}(r) = \mp \sqrt{ \frac{ \ell(\ell+1)}{2}} \frac{1}{r}
	\left(V_{\ell m_A} \mp U_{\ell m_A} \right) .
\ee

\section{The Equilibrium Magnetic Field}\label{equilmag}

A static magnetic field which is symmetric about an axis can be written in
the general form
\st 
\be
\label{eqB1}
\B(r,\theta',\phi') = \sum_{\ell} \sqrt{ \frac{4\pi}{2\ell+1}}
	\left( \beta_{1,\ell}(r) Y_{\ell0}(\theta',\phi') \unit_{\hat r}
	- r \beta_{2,\ell}(r) \nab' Y_{\ell0}(\theta',\phi') 
	-i \beta_{3,\ell}(r) \opL' Y_{\ell0}(\theta',\phi') \right),
\ee
where $\theta'$ and $\phi'$ are angular coordinates measured with respect
to the magnetic field's axis of symmetry and $\nab'$ and $\opL'$ are 
angular operators with respect to the magnetic field's symmetry axis. 
The expansion of the general magnetic field in the spin-weighted basis
is
\st
\be
B(r,\theta',\phi') = 
	 B'_{0}(r,\theta',\phi') \unit_0  +
	 B'_{+}(r,\theta',\phi') \unit_-  +
	 B'_{-}(r,\theta',\phi') \unit_+  ,
\ee
where each component has the expansion in spin-weighted spherical
harmonics 
\st\be
{B'}_{s} = \sum_\ell {b'}_{s}^{\ell}(r) \;_s\!Y_{\ell 0}(\theta',\phi'),
\ee
and the set of functions ${b'}_s^{\ell}(r)$ are given by
\st
\be
{b'}_{0}^{\ell} = \sqrt{\frac{4\pi}{2\ell+1}} \beta_{1,\ell}
\quad \hbox{and} \quad
{b'}_{\pm}^{\ell} = \pm \sqrt{\frac{4\pi}{2\ell+1}}
		\sqrt{\frac{\ell(\ell+1)}{2}} 
		(\beta_{2,\ell} \pm i \beta_{3,\ell}).
\ee

In general, the axis of symmetry of the magnetic field will not coincide
with the star's spin axis. Suppose that the magnetic field is tilted by
an angle $\alpha$ from the star's spin axis. If the angles $\theta$ and
$\phi$ are measured with respect to the spin axis, the spin-weighted 
spherical harmonics in the two coordinate systems are related by
\st
\be
 _{s}\!Y_{\ell k}(\theta',\phi') = 
	\sum_m  d^{\ell}_{mk}(\alpha) \; _{s}\!Y_{\ell m}(\theta,\phi).
\ee
As a result, the magnetic field can also be written in terms of 
the spin-axis coordinate system
\st
\begin{mathletters}
\bea
B(r,\theta,\phi) &=& 
	 B_{0}(r,\theta,\phi) \unit_0  \;+\;
	 B_{+}(r,\theta,\phi) \unit_-  \;+\;
	 B_{-}(r,\theta,\phi) \unit_+  \label{Bb}\\
&& B_{s} = \sum_{\ell,m} b_{s}^{\ell,m}(r) \;_s\!Y_{\ell m}(\theta,\phi)\\
&& b_{s}^{\ell,m}(r) = {b'}_{s}^{\ell}(r) d^{\ell}_{m0}(\alpha).
\eea
\end{mathletters}		
In order for the magnetic field to be real, the coefficients must
obey the relations
\st
\begin{mathletters}
\bea
b_{0}^{\ell,m*}(r) &=& b_{0}^{\ell,m}(r) \\
b_{+}^{\ell,m*}(r) &=& - b_{-}^{\ell,m}(r)
\eea
\end{mathletters}
where $*$ denotes complex conjugation.

The exterior magnetic field of most stars is approximately
that of a dipole, which has the form
$\B(r',\theta',\phi') = - B_p R^3 \sqrt{ {4\pi}/{3}}\; \nab' 
(Y_{1\;0}(\theta',\phi')/r'^2)$
in the coordinate system aligned with the magnetic field's symmetry axis. 
In this case the exterior magnetic is described by coefficients
$\beta_1(r)/2 = \beta_2(r) = B_p{R^3}/{r^3}$ and $\beta_3(r)=0$. 
(We will now drop the subscript $\ell$ for the $\beta$ coefficients
since we will only be considering the dipole case $\ell=1$.)
In the case of a dipole field,
the only coefficients of the spin matrix
$d^1(\alpha)$ of interest are 
\st
\be
d^{1}_{10}(\alpha) = \frac{1}{\sqrt{2}} \sin \alpha, \quad
d^{1}_{00}(\alpha) = \cos \alpha, \quad
d^{1}_{-10}(\alpha)= -\frac{1}{\sqrt{2}} \sin \alpha.
\label{d1}
\ee
In the coordinate system aligned with the star's spin axis,
the exterior magnetic field is then described by the functions
\st\be \label{exter}
\frac{1}{2} b^{1m}_0(r) = b^{1m}_+(r) = - b^{1m}_-(r)
	= \sqrt{ \frac{4\pi}{3}} d^{1}_{m0}(\alpha)  B_p \frac{R^3}{r^3}.
\ee

In order to compute the eigenmodes of a star with an dipole magnetic
field exterior to the star, the
magnetic field inside of the star must be specified. 
In this paper we will only consider interior fields which are
dipole (i.e. $\ell = 1$). General formulae for
the perturbations of a star with any interior dipole field will 
be presented in the following section. As examples we will 
consider two simple interior magnetic fields which can be
matched to the dipole exterior (in the sense that
the normal component of the magnetic field at the star's 
surface is continuous). Both of these examples are
force-free fields which obey 
\st
\be
\label{ff}
\nab \times \B = \mu \B,
\ee
where $\mu$ is a constant with units of inverse length.
Force-free magnetic fields have the simple feature that they
do not distort the star's equilibrium fluid configuration.

The first example is the simplest possible interior field,
the uniform magnetic field, corresponding
to $\mu=0$ in equation (\ref{ff}).
The coefficients appearing
in equation (\ref{eqB1}) are 
\st\be
\label{unibeta}
\beta_1=-\beta_2=B_{in}, \quad \beta_3=0,
\ee
where $B_{in}$ is the magnetic field at the centre of the star.
Since the radial component of the magnetic field must be 
continuous at the surface of the star, $B_{in} = 2 B_p$.
The 
magnetic field coefficients in the coordinate system aligned with
the spin axis are
\st\be
b_0^{1,{m}}(r) = b_-^{1,{m}}(r) = -b_+^{1,{m}}(r)
	= \sqrt{\frac{4\pi}{3}}B_{in} d^{1}_{{m}0}(\alpha).
\label{unib}
\ee

The second example corresponds to the force-free condition
(\ref{ff}) with $\mu$ nonzero discussed by \citet{FP66}. 
The magnetic field coefficients are 
\st\be
\beta_1(r)=3B_{in}\frac{j_1(\mu r)}{\mu r},~~
\beta_2(r)=-\frac32 B_{in}\frac{1}{\mu}\lp\frac{d}{dr}+\frac{1}{r}\rp j_1(\mu r),~~
\beta_3=-\frac32 B_{in}j_1(\mu r),
\ee 
where
$j_1(\mu r)$ 
is a spherical Bessel function.
Continuity of the normal component of the magnetic field
at the surface of star requires that the constant
$\mu$ be given by the solution of the equation 
\st
\be
\label{j1eq}
j_1(\mu R) = \frac23\frac{B_p}{B_{in}}\mu R,
\ee
which can be solved numerically once the ratio ${B_p}/B_{in}$ is specified.
The solution to equation (\ref{j1eq}) is shown in Figure 1, where the 
dimensionless product $\mu R$ is plotted versus the ratio ${B_p}/B_{in}$.
The force-free magnetic field then corresponds to family of fields
parametrised by the ratio of magnetic field at the pole to the 
magnetic field at the centre of the star. This ratio of magnetic 
fields can have values in the range $0 < {B_p}/B_{in} \le 0.5$, as
shown in Figure 1.

\section{Magnetically Modified r-modes}
\label{rmodes}

We now turn to the calculation of the modification of the
r-modes due to  dipole magnetic field. 
In the slow-rotation approximation, the r-modes of non-magnetic
incompressible stars have the form
\st
\begin{mathletters}
\bea
&& \omega_{\ell_A m_A} = - \frac{2m_A\Omega}{\ell_A(\ell_A+1)}\\
&& W_{\ell_A m_A} = 0, \quad V_{\ell_A m_A} = 0 \\
&& U_{\ell_A m_A} = \delta_{\ell_A |m_A|} R^2
\sqrt{ \frac{2\pi (2\ell_A+3)(\ell_A+1)}{15 \ell_A}} 
\left( \frac{r}{R} \right)^{\ell_A+1}
\\
&& \epsilon_{\ell_A m_A} =  T = \frac{1}{5} M R^2 \Omega^2,
\eea
\end{mathletters}        
where the functions $W_{\ell_A m_A}, V_{\ell_A m_A}$ and 
$U_{\ell_A m_A}$ are defined in the expansion (\ref{genformmodes}).
Each r-mode solution is identified by a distinct quantum number
$m_A$. At lowest order in angular velocity, the only value of $\ell_A$
allowed is $\ell_A=|m_A|$. The spin-weighted decomposition of the
$A$th r-mode is
\st\be
\xi_A(x) = f^A_{0}(r) \;_0\!Y_{\ell_Am_A}  \unit_0 + 
f^A_{+}(r) \;_+\!Y_{\ell_Am_A}  \unit_- + 
f^A_{-}(r) \;_-\!Y_{\ell_Am_A}  \unit_+ ,
\ee
where the functions $f_s^A(r)$ are given by
\st\be
f^A_0(r) = 0,\quad 
f^A_+ = f^A_- = R (\ell_A+1) 
\sqrt{ \frac{\pi (2\ell_A+3)}{15}}
   \left(\frac{r}{R}\right)^{\ell_A}.
\label{f:rmode}
\ee

In order to compute the r-mode frequencies in the presence of a magnetic
field, we must calculate the values of the magnetic coupling coefficients 
appearing in equation (\ref{KAD}). The first coupling coefficient $\kappa^{(1)}_{AD}(r)$
depends on the perturbed magnetic field. The equations for the perturbed magnetic
field were derived in the appendix for general magnetic fields and general 
fluid perturbations. In this section we will present the form of the perturbed
magnetic field for r-modes in the presence of a dipole magnetic field.
In general, the perturbed magnetic field has the expansion
\st\be
\label{deltaB1}
\delta \B_A(r,\theta,\phi) = \sum_{\lambda,\mu}\left(
	 \delta B_{A,0}^{\lambda,\mu}(r) \;_0\!Y^*_{\lambda \mu}\unit_0  \;+\;
	 \delta B_{A,+}^{\lambda,\mu}(r) \;_-\!Y^*_{\lambda \mu}\unit_-  \;+\;
	 \delta B_{A,-}^{\lambda,\mu}(r) \;_+\!Y^*_{\lambda \mu}\unit_+\right) .
\ee
In this equation, the summation is over all $\lambda$ obeying 
the triangle inequality $\ell_A -1 \le \lambda \le \ell_A +1$ 
 and all $\mu$ obeying $\mu = -\ell_A -{m}$. In the 
case of an r-mode and a dipole magnetic field, the ``0'' component appearing
in equation (\ref{deltaB1}) is given by
\st\be
\delta B^{\lambda,\mu}_{A,0} = \sum_{{m}=-1}^{+1} 
	C(\ell_A,1,\lambda) 
	\left( \begin{array}{ccc}
		\lambda & \ell_A & 1\\
		\mu	& m_A & {m}
		\end{array}\right) 
	\left( \begin{array}{ccc}
		\lambda & \ell_A & 1\\
		0	& -1 & 1
		\end{array}\right) \frac{1}{r} b^{1,{m}}_0 f^A_+(r)
	\left( 1 + (-1)^{\lambda+\ell_A} \right),
\ee
which is only nonzero when $\mu = -(m_A + {m})$ and 
$\lambda = \ell_A$ and the constant $C(\ell_A,1,\lambda)$ is given
by 
\st\be
C(\ell_A, {\ell},\lambda) = (-1)^{\ell_A+\lambda+ {\ell}} 
\sqrt{ \frac
{(2\ell_A+1)(2{\ell}+1)(2\lambda+1)}{4\pi}}.
\ee
The $\pm$ components appearing in (\ref{deltaB1}) are
\st
\begin{mathletters}
\bea
\delta B^{\lambda,\mu}_{A,+}& =& (-1)^{\lambda+\ell_A+1} 
{\delta B^{\lambda,\mu}_{A,-}}^*\label{sym}  \\
&=&\sum_{{m}=-1}^{+1} 
	C(\ell_A,1,\lambda) 
	\left( \begin{array}{ccc}
		\lambda & \ell_A & 1\\
		\mu	& m_A & {m}
		\end{array}\right)  \frac{1}{r} f^A_+(r)  
\times \label{deltab+}\\
&& \left[
	\lp (\ell_A-1) b_{0}^{{1,m}}(r)  
          - b_{+}^{{1,m}}(r)\rp
               \left( \begin{array}{ccc}
		\lambda & \ell_A & 1\\
		-1	& 1 & 0
		\end{array}\right) 
	+ \sqrt{ \frac{\ell_A(\ell_A+1)}{2}}   b_{+}^{{1,m}}(r)
\left( \begin{array}{ccc}
		\lambda & \ell_A & 1\\
		-1	& 0 & 1
		\end{array}\right) \right.\no\\
&&\left.\hspace{5cm}
-  \sqrt{ \frac{(\ell_A-1)(\ell_A+2)}{2}}  b_{-}^{{1,m}}(r)
\left( \begin{array}{ccc}
		\lambda & \ell_A & 1\\
		-1	& 2 & -1
		\end{array}\right)
\right].\nonumber 
\eea
\end{mathletters}
In equation (\ref{deltab+}) $\lambda$ takes on values $\ell_A, \ell_A\pm1$. 
Evaluating the terms appearing in square brackets in equation (\ref{deltab+}),
the individual terms for the allowed values of $\lambda$ are 
\st
\begin{mathletters}
\bea
\delta B^{\ell_A,\mu}_{A,+}&=&
(-1)^{\ell_A+1}  
\sqrt{ \frac{3 (2\ell_A+1)}{4\pi\ell_A(\ell_A+1)}} \frac{f^A_+(r)}{r}\\ 
&&\times
\sum_{{m}=-1}^{+1} \ls 
	(\ell_A-1) b^{1,{m}}_0 
	- (b^{1,{m}}_+ + b^{1,{m}*}_+)
             -\frac12\ell_A(\ell_A+1)( b^{1,{m}}_+ - b^{1,{m}*}_+)\rs 
	\left( \begin{array}{ccc}
		\ell_A & \ell_A & 1\\
		-(\ell_A+{m})	& \ell_A & {m}
		\end{array}\right)  
\no\\
\delta B^{\ell_A-1,\mu}_{A,+}&=&
(-1)^{\ell_A+1} 
\sqrt{ \frac{3(\ell_A-1) (\ell_A+1)}{4\pi\ell_A}}
\frac{f^A_+(r)}{r}\\
&&\times
\sum_{{m}=-1}^{+1} \ls 
	(\ell_A-1) b^{1,{m}}_0 
             -\frac12(\ell_A+2)( b^{1,{m}}_+ + b^{1,{m}*}_+)\rs 
	\left( \begin{array}{ccc}
		\ell_A-1 & \ell_A & 1\\
		-(\ell_A+{m})	& \ell_A & {m}
		\end{array}\right)  
\no\\
\delta B^{\ell_A+1,\mu}_{A,+}&=& 
(-1)^{\ell_A} 
\sqrt{ \frac{3\ell_A (\ell_A+2)}{4\pi(\ell_A+1)}} 
\frac{f^A_+(r)}{r}\\
&&\times
\sum_{{m}=-1}^{+1} \ls 
	(\ell_A-1) b^{1,{m}}_0 
             +\frac12(\ell_A-1)( b^{1,{m}}_+ + b^{1,{m}*}_+)\rs 
	\left( \begin{array}{ccc}
		\ell_A+1 & \ell_A & 1\\
		-(\ell_A+{m})	& \ell_A & {m}
		\end{array}\right).
\no
\eea
\end{mathletters}

The first coupling coefficient $\kappa^{(1)}_{AD}$ is given by equation (\ref{kappa1}).
The first summation appearing in equation (\ref{kappa1})
reduces to 
\st\be
\sum_{\lambda,\mu}\delta B^{*\lambda,\mu}_{A,0} \delta B^{\lambda,\mu}_{D,0}
 = 4 \delta_{AD} C^2(\ell_A,1,\ell_A) \sum_{{m}=-1}^{+1} 
	\left(\frac{1}{r} b^{1,{m}}_0 f^A_+(r)\right)^2
	\left( \begin{array}{ccc}
		\ell_A & \ell_A & 1\\
		\mu	& m_A & {m}
		\end{array}\right)^2
	\left( \begin{array}{ccc}
		\ell_A & \ell_A & 1\\
		0	& -1 & 1
		\end{array}\right)^2.
\label{term1}
\ee
Making use of the known values of the Wigner 3-j symbols \citep{Edm74},
equation (\ref{term1}) reduces to 
\st\be
\sum_{\lambda,\mu}\delta B^{*\lambda,\mu}_{A,0} \delta B^{\lambda,\mu}_{D,0}
 = \frac{2}{\ell_A+1} \delta_{AD} \left(\frac{1}{r}\beta_1f^A_+(r)\right)^2 
	\left( \ell_A \cos^2\alpha + \frac12 \sin^2\alpha\right).
\ee
Due to the symmetry property (\ref{sym}), the final two summations appearing in 
(\ref{kappa1}) are
\st\be
\sum_{\lambda,\mu}\left(\delta B^{*\lambda,\mu}_{A,+} \delta B^{\lambda,\mu}_{D,+}
+ \delta B^{*\lambda,\mu}_{A,-} \delta B^{\lambda,\mu}_{D,-} \right)
= \sum_{\lambda,\mu}\delta B^{*\lambda,\mu}_{A,+} \delta B^{\lambda,\mu}_{D,+}
+ (-1)^{\ell_A+\ell_D}\delta B^{\lambda,\mu}_{A,+} \delta B^{*\lambda,\mu}_{D,+}.
\ee
Since the triangle inequalities $\ell_A-1\le \lambda \le \ell_A+1$
and $\ell_D-1\le \lambda \le \ell_D+1$ must both be satisfied, only modes satisfying 
$\ell_D = \ell_A, \ell_A\pm1, \ell_A\pm2$ have non-zero magnetic coupling. 
As a result the only nonzero entries in the coupling coefficient
matric $\kappa_{AD}$ will be those coupling the modes 
satisfying the triangle inequality.

The general equation for the diagonal elements of the 
coupling coefficient matrix $\kappa^{(1)}_{AA}(r)$,
for any interior dipole magnetic field is
\st
\bea
\kappa^{(1)}_{AA}(r) &=& 2 \left( \frac{f_+(r)}{r} \right)^2 \times
 \left[ \frac{1}{\ell_A+1}(\beta_1)^2\left( \ell_A \cos^2\alpha + \frac12 \sin^2\alpha\right) \right.
\label{genkap1}
\\
&& \qquad 
 + \frac{1}{\ell_A(\ell_A+1)^2}
		\left( \left((\ell_A-1)\beta_1 - 2\beta_2 \right)^2 
		+ \left( \beta_3 \right)^2 \right) 
		 \left( \ell_A \cos^2\alpha + \frac12 \sin^2\alpha\right)
\nonumber \\
&&	
        \qquad + \frac{ (\ell_A-1)(\ell_A+1)}{2\ell_A (2\ell_A+1)}
	\left( (\ell_A-1)\beta_1 - (\ell_A+2) \beta_2 \right)^2
	\sin^2\alpha
\nonumber \\
&&	\left.
	\qquad + \frac{\ell_A (\ell_A+2)(\ell_A-1)^2 }{(2\ell_A+3)(\ell_A+1)^2} 
	\left(\beta_1+\beta_2\right)^2 
	\left( \cos^2\alpha + \frac{(2\ell_A^2 + 3\ell_A + 2)}{2 (2\ell_A+1)}
	\right) \right]
\nonumber 
\eea
Similarly, the function $\kappa^{(2)}_{AA}$ has the value
\st
\bea
\kappa^{(2)}_{AA}(r) &=& \left( \frac{f_+(r)}{r} \right)^2 
	\frac{1}{(\ell_A+1)(2\ell_A+1)} 
\times \label{genkap2}
\\
&& \qquad \left( 2 \beta_1\left( \frac{d}{dr} (r\beta_2) + \beta_1\right)
	- 2 \left(\beta_3\right)^2 \right)
	\left(\ell_A \cos^2\alpha + \frac12\sin^2\alpha\right)
\nonumber 
\eea
The frequency correction can now be computed for any interior 
dipole magnetic field, once the functions $\beta_1(r), \beta_2(r)$
and $\beta_3(r)$ which describe the equilibrium 
magnetic field (see equation (\ref{eqB1}))
have been determined.


\subsection{Frequency Correction for a Uniform Magnetic Field}

The simplest example of an interior magnetic field is the uniform
magnetic field which has $\beta_1(r) = - \beta_2(r) = B_{in}$
and $\beta_3(r) = 0$. 
Since  the magnetic field is
uniform, it satisfies the equation $\nabla \times \B = 0$. As a result,
the term $\kappa_{AD}^{(2)}$ vanishes. The term $\kappa_{AD}^{(3)}$
does not vanish, but its contribution to the complete coupling coefficient
$\kappa_{AD}$ is a factor of $\Omega^2R^2/c^2$ smaller than the 
contribution from $\kappa_{AD}^{(1)}$, so we neglect it. 
The magnetic coupling coefficients between different r-modes can now
be computed. The only nonvanishing coefficient is the
 self-coupling term $\kappa_{AA}$ given by equation (\ref{genkap1}). A  
straight-forward calculation yields
\st\be
\kappa_{AA} = - \frac15 (\ell_A+1)(2\ell_A+3) \left[ 1
+ \frac12 (\ell_A(\ell_A+1)-3) \sin^2\alpha\right].
\label{final1}
\ee
Since the matrix $\kappa$ is diagonal for the case of 
coupled r-modes, the frequencies $\sigma_A$ of the magnetically
modified r-modes are simply
\st\be
\sigma_{A} = \omega_A \left(1 + \frac{\cal{M}}{2T} |\kappa_{AA}|\right).
\label{final2}
\ee

The final equations (\ref{final1}) and (\ref{final2})
 for the frequency correction for the r-modes is 
surprisingly simple. However, this simplicity is a result of the
simplicity of the background magnetic field. Since we assume that
the field is homogeneous, terms involving derivatives of 
the magnetic field vanish, with the net result that the matrix
$\kappa_{AD}$ is diagonal. 

In this simple example, we have not considered couplings between
r-modes and other families of modes, such as the f-modes or
inertial modes. By neglecting these couplings, we are 
neglecting off-diagonal terms in the matrix $\kappa_{AD}$, which
is only valid in the approximation that the energy in the 
magnetic field is small compared to the star's rotational
and gravitational potential energies. For this reason equation
(\ref{final2}) is the frequency correction to the r-modes only
in the weak-field approximation. We will compute the couplings
to other modes elsewhere, which will allow the
calculation of the modes of rotating stars with arbitrarily
large magnetic field.

\subsection{Frequency Corrections for a force-free Magnetic Field}\label{uniformB}

The second model for the interior magnetic field which we 
have examined is the force-free magnetic field, which is described
by the functions $\beta_1(r),
\beta_2(r)$ and $\beta_3(r)$
given by equation (\ref{unibeta}). All three functions are proportional
to $B_{in}$, which is the magnetic field strength at the centre of the
star, which is not a known quantity.
Typically, only the exterior dipole
magnetic field at the pole of the star, $B_p$, 
can be measured, and the interior
magnetic field is unknown. When the interior field is modeled by
a force-free field, the ratio of $B_{p}/B_{in}$ is a free parameter.
The smallest ratio
of interior to exterior magnetic fields allowed in the force-free
magnetic field model is $B_{p}/B_{in} = 0.5$, which corresponds to 
the uniform magnetic field (see Figure 1).

Due to the toroidal symmetry 
of the fluid displacement vector, $\kappa^{(2)}_{AA}$ vanishes
for the force-free magnetic field
so that only $\kappa^{(1)}_{AA}$ needs to be computed. This computation
can be easily done numerically and the results are shown in
Table 2 and Figure 2. The frequency correction depends on the strength
of the internal magnetic field and the angle between the spin axis and
magnetic field axis. In order to show the dependence on both of these
parameters, we have written the correction term in the form
$\kappa_{AA} = a_{AA} + b_{AA} \sin^2\alpha$, where $a_{AA}$ 
and $ b_{AA}$ depend only on the ratio of external to internal magnetic field
$B_p/B_{in}$.  
In Table 2 we show the values of the magnetic
frequency correction terms for various values of $\ell_A$ and for the specific
value of magnetic field ratio $B_p/B_{in}=0.1$. It can be seen that 
the coupling coefficients grow with increasing values of the spherical
harmonic index $\ell_A$. This property agrees with what is seen in 
the case of the uniform magnetic field.

In Figure 2 we show the values
of the frequency correction for the case $\ell_A=m_A=2$ for the complete
 range of allowed interior magnetic field strengths. 
Figure 2 shows
that the magnitude of the 
magnetic coupling coefficients decreases as the interior 
magnetic field is increased, which seems counter-intuitive. However, 
in order to find the corrected frequency in
equation (\ref{final2}), the coupling coefficient 
is multiplied by $\cal{M}$, the energy of the internal magnetic field.
Now ${\cal{M}} \propto B_{in}^2$, and the internal magnetic field is
not known: all that we can measure is the external dipole magnetic
field strength $B_p$. Once $B_p$ is known, the fraction $B_p/B_{in}$
is a free parameter of the theory. Therefore, the true scaling of
the frequency correction due to magnetic fields is
$(B_{in}/B_{p})^2 \kappa_{AA}$. Once this is taken into account,
we see that the magnitude of the frequency correction increases
with increasing interior magnetic field strength, as would 
be expected.

\section{Conclusions}\label{conclusions}

In this paper we have presented a formalism for computing
the oscillation modes of magnetic rotating stars. This method
is based on the phase-space mode decomposition introduced by
\citet{Sch01} and makes use of spin-weighted spherical harmonics.
As a simple example, we have computed the magnetic corrections
to the r-modes of an incompressible star with a uniform magnetic
field in the weak-magnetic field limit. In this limit, the
spatial form of the r-mode velocity field is unchanged by the
presence of a magnetic field. In the case of a uniform interior magnetic
field matched to an exterior dipole field, the resulting 
formula for the frequency has the simple form presented in equations
(\ref{final1}) and (\ref{final2}). Qualitatively similar results 
were found for the force-free magnetic field. 
The presence of a magnetic field
increases the absolute value of the rotating frame frequency, so that 
the mode will be counter-rotating {\em faster} than if the star
had no magnetic field. Adding a magnetic field to the star increases the
restoring force acting on the fluid and increases the oscillation frequency,
just as increasing the spring constant of a spring causes a spring to 
undergo faster oscillations.  Mathematically, the increase in frequency
results from the fact that the diagonal entries in the matrix $\kappa$
are negative. Since these diagonal entries correspond to the work
done by the magnetic field on the fluid, the negative sign denotes the
fact that the magnetic field opposes the fluid motion. Hence, the
increase in frequency which we have found agrees with basic physical
intuition. 

Although we have only presented detailed results for two types of
internal dipole field, the equations presented in section~\ref{rmodes} can be used to 
determine the frequency correction for any other internal dipole
field. Once the functions $\beta_1(r) - \beta_3(r)$ describing the
equilibrium magnetic field have been found, it is only necessary
to substitute them into equations (\ref{genkap1}) and (\ref{genkap2}) 
and perform the integration over
volume in equation (\ref{KAD}).

As long as the magnetic field energy is small
compared to the star's rotational energy, the CFS 
gravitational-radiation-driven instability 
criterion \cite{FS78b} will be unchanged. As a result, the
magnetic field has a slight stabilizing effect on the
r-modes, since it is now harder for the mode to be dragged 
forward by the star's rotation. This agrees with the 
intuitive notion that magnetic fields tend to oppose 
fluid motion. However, in order for 
the mode to be stable to the CFS mechanism, it would be
necessary for the magnetic field's energy to be of the
same order of magnitude as the rotational energy, 
which corresponds to a regime where our present results
do no hold. Such a large field would presumably alter the 
stability criterion \citep{FS78b}  as well.

Although our calculations for the r-modes of magnetic stars
is only valid in the weak-field limit, we note that a young
neutron star with spin period and magnetic field similar to
the Crab's will have ${\cal{M}}/T$ of the order $10^{-9}$
well within the regime of the weak-field limit. Since 
we do not include a crust, the applicability of our equations
is mainly to young neutron stars with $T > 10^{10} K$ or 
other stars without a solid layer.

Our calculations are purely linear, so the present work 
does not shed any light on the nonlinear evolution of
an unstable mode of a magnetic star which has been
investigated by \citet{Rez00}. However, 
the frequency change due to the
magnetic field which we have calculated  will alter the kinematic
drift found by \citet{Rez00}, although we doubt that these
changes would alter their qualitative results. 
It would be interesting to see how the magnetically modified
r-modes presented in this paper affect the evolution of spin
and magnetic field in young neutron stars.

\acknowledgments

This research was supported by the Natural
Sciences and Engineering Research Council of Canada.
We would like to thank Roy Maartens, Draza Markovi\'{c} 
and Luciano Rezzolla for useful discussions and comments
about this paper.


\appendix

\section{Derivation of the Magnetic Coupling Coefficients for General Magnetic
Fields}

In this appendix we give details about the calculation of the magnetic 
coupling coefficients which appear in equation (\ref{KAD}). 
The first coupling coefficient $\kappa^{(1)}_{AD}$ depends on the 
perturbed magnetic field, defined by
$\delta \B_A = \nabla \times (\xi_A \times B)$, where $B$ is the
equilibrium magnetic field and $\xi_A$ is the fluid displacement
vector. This definition of the perturbed magnetic field can
be rewritten in the form
\st\be\label{dB-sw-1}
\delta B_A^a=(\xi_A^a B^b-\xi_A^b B^a)_{;b},
\ee
where the subscripted semicolon denotes covariant differentiation,
and lower case Latin subscripts and superscripts are vector indices.
The components of the perturbed magnetic field in the 
basis $(\unit_0,\unit_+,\unit_-)$ can be found by 
substituting the expansions (\ref{xispin}) and (\ref{Bb})
into equation (\ref{dB-sw-1}).
The radial component of the perturbed magnetic field is given by
\st\bea\label{dB-0-comp}
\delta B_A^0 &=& \delta B_A\cdot \unit_0  ={\xi_A^0}_{~;b} B^b + \xi_A^0 B^b_{~;b} -
 {\xi_A^b}_{~;b} B^0 -\xi_A^b B^0_{~;b}
\no\\
&=& {\xi_A^0}_{~;+} B^+ + {\xi_A^0}_{~;-} B^-
+ \xi_A^0 B^+_{~;+} +  \xi_A^0 B^-_{~;-}
 - {\xi_A^+}_{~;+} B^0 -{\xi_A^-}_{~;-} B^0
- \xi_A^+ B^0_{~;+} - \xi_A^- B^0_{~;-}\,.
\eea
Note that although it is true that $B^b_{~;b}=0$, we gain an 
equation with a more symmetric form by not explicitly setting
the divergence of the magnetic field to zero. 
Since each component of the equilibrium magnetic field and the fluid 
displacement vector is proportional to a spin-spherical
harmonic, the perturbed magnetic field components can be written 
in the compact form
\st
\be
\delta B_{A,s}(r,\theta,\phi)
	= \sum_{\ell_A} \sum_{{\ell},{m}} \sum_{s_A+{s} = s}
	\zetalamb_{s,s_A,{s}}(r) \;_{s_A}Y_{\ell_A m_A} 
		\;_{{s}}Y_{{\ell}{m}}
\label{dBspin}
\ee
where $s$ takes on values of 0 and $\pm 1$, and the radial 
functions $\zetalamb_{s,s_A,{s}}(r)$ will be determined shortly.
In equation (\ref{dBspin}), the integers $\ell_A, m_A, s_A$ are the fluid perturbation's
quantum numbers, ${\ell}, {m}, {s}$ are the equilibrium
magnetic field's quantum numbers and the symbol $\Lambda$ represents the set of
quantum numbers $({\ell}, {m})$ describing the magnetic field.

A detailed formalism for taking covariant derivatives of 
vectors using spherical coordinates has been given in 
section VI of \citet{Sch01}. Their equations (6.15) and (6.16)
summarise the rules for taking covariant derivatives in 
the spin-weighted basis. (The only difference in notation between
the present paper and that of \citet{Sch01} is the symbols used
for the basis vectors. \citet{Sch01} use the basis $\{ \mathbf l,
{\mathbf m}, \bar{\mathbf m} \}$ which are related to our basis by
$ \mathbf l = \unit_0$, $\mathbf m = \unit_+$ and $\mathbf{\bar{m}}
= \unit_-$.)
When these covariant differentiation rules are used, we find
that the $s=0$ terms of $\zetalamb$ appearing in (\ref{dBspin}) are
\st
\begin{mathletters}
\bea
\zetalamb_{0,0,0}(r) &=& 	
	\sqrt{ \frac{{\ell}({\ell} +1)}{2}}
	\frac{\fzA}{r} (\bplamb - \bmlamb)
	- \sqrt{ \frac{{\ell_A}({\ell_A} +1)}{2}}
	\frac{\bzlamb}{r} (\fpA - \fmA)
	\label{zeta:a}\\
\zetalamb_{0,1,-1}(r) &=& - \left( 
	 \sqrt{ \frac{{\ell}({\ell} +1)}{2}}
	\frac{\fpA \bzlamb}{r} 
	+ \sqrt{ \frac{{\ell_A}({\ell_A} +1)}{2}}
	\frac{\fzA \bmlamb}{r} \right)
	\label{zeta:b}\\
\zetalamb_{0,-1,1}(r) &=&	
	\sqrt{ \frac{{\ell}({\ell} +1)}{2}}
	\frac{\fmA \bzlamb}{r} 
	+ \sqrt{ \frac{{\ell_A}({\ell_A} +1)}{2}}
	\frac{\fzA \bplamb}{r}. 
\label{zeta:c}
\eea
\end{mathletters}

Similar manipulations give us the $\pm$ components of the
perturbed magnetic field.
In particular, the $+$ component of $\delta B$ is
\st\bea\label{dB-+-comp}
\delta B_+ =
\delta B^- = \delta B\cdot \unit_+ = &&
	\xi^-_{~;b} B^b + \xi^- B^b_{~;b} - \xi^b_{~;b} B^- -\xi^b B^-_{~;b}
\no\\
=&& \xi^-_{~;+} B^+ + \xi^-_{~;0} B^0
+ \xi^- B^+_{~;+} +  \xi^- B^0_{~;0}\no\\
&& - \xi^+_{~;+} B^- -\xi^0_{~;0} B^-
- \xi^+ B^-_{~;+} - \xi^0 B^-_{~;0}\,.
\eea
Substitution of the covariant differentiation rules into 
(\ref{dB-+-comp}) and comparing with equation (\ref{dBspin})
results in 
\st
\begin{mathletters}
\bea
\zetalamb_{1,1,0}(r) &=& \frac{1}{r} \frac{d}{dr}\left(
	r \bzlamb \fpA\right) 
	- \sqrt{ \frac{{\ell}({\ell} +1)}{2}}
	\frac{ \bmlamb \fpA}{r} \label{zeta:d}
\\
\zetalamb_{1,0,1}(r) &=& -\frac{1}{r} \frac{d}{dr}\left(
	r \bplamb \fzA\right) 
	+ \sqrt{ \frac{{\ell}_A({\ell}_A +1)}{2}}
	\frac{ \bplamb \fmA}{r}
 \\
\zetalamb_{1,2,-1}(r)&=& - \sqrt{ \frac{({\ell}_A-1)({\ell}_A +2)}{2}}
	\frac{ \bmlamb \fpA}{r}
\\
\zetalamb_{1,-1,2}(r)&=& \sqrt{ \frac{({\ell}-1)({\ell} +2)}{2}}
	\frac{ \bplamb \fmA}{r}.
\label{zeta:g}
\eea
\end{mathletters}
In the case of a purely dipole magnetic field, ${\ell}=1$ 
and the term $\zetalamb_{1,-1,2}(r)$ vanishes.
Similarly, the $s=-1$ components of $\zeta$ are
\st
\begin{mathletters}
\bea
\zetalamb_{-1,-1,0}(r) &=& \frac{1}{r} \frac{d}{dr}\left(
	r \bzlamb \fmA\right) 
	+ \sqrt{ \frac{{\ell}({\ell} +1)}{2}}
	\frac{ \bplamb \fmA}{r} \label{zeta:h}
\\
\zetalamb_{-1,0,-1}(r) &=& -\frac{1}{r} \frac{d}{dr}\left(
	r \bmlamb \fzA\right) 
	- \sqrt{ \frac{{\ell}_A({\ell}_A +1)}{2}}
	\frac{ \bmlamb \fpA}{r}
 \\
\zetalamb_{-1,-2,1}(r)&=&  \sqrt{ \frac{({\ell}_A-1)({\ell}_A +2)}{2}}
	\frac{ \bplamb \fmA}{r}
\\
\zetalamb_{-1,1,-2}(r)&=& -\sqrt{ \frac{({\ell}-1)({\ell} +2)}{2}}
	\frac{ \bmlamb \fpA}{r}.
\label{zeta:k}
\eea
\end{mathletters}

In equation (\ref{dBspin}), the components of perturbed magnetic field were 
written as products of spin-spherical harmonics. This equation can be simplified
so that each term is proportional to only one spin-spherical harmonic,
\st\be
\delta \B_{A,s}(r,\theta,\phi) = \sum_{\lambda,\mu}
	 \delta B_{A,s}^{\lambda,\mu}(r) \;_s\!Y^*_{\lambda \mu},
\ee
by using the combination formula 
\st\be
\;_{s}\!Y_{\ell m}(\theta,\phi) \;_{t}\!Y_{k n}(\theta,\phi) 
  = \sum_{\lambda,\mu,\sigma} \sqrt{ \frac{(2\ell+1)(2k+1)(2\lambda+1)}{4\pi}} 
	\left( \begin{array}{ccc}
	\ell&k&\lambda\\
	-s&-t&-\sigma \end{array}\right)
	\left( \begin{array}{ccc}
	\ell&k&\lambda\\
	m&n&\mu \end{array}\right) \;_{\sigma}\!Y^*_{\lambda \mu}(\theta,\phi)
\label{combo}
\ee
where the summation is over all values of $\lambda$ satisfying the triangle inequality
$|\ell-k|\le \lambda \le \ell+k$, 
$\mu$ satisfying $\mu = -(m+n)$, $\sigma$ satisfying $\sigma=-(s+t)$ and 
$\left( \begin{array}{ccc}
\ell&k&\lambda\\
-s&-t&-\sigma \end{array}\right) $ is a Wigner 3-j symbol \citep{Edm74}. 
The radial expansion functions $\delta B_{A,s}^{\lambda,\mu}(r)$
are given by
\st\be
\delta B_{A,s}^{\lambda,\mu}(r) = \sum_{{m}=-1}^{+1}
	C(\ell_A,1,\lambda)  
	\left( 
	\begin{array}{ccc}
	\lambda&\ell_A&1\\
	\mu&m_A&{m}
	\end{array}\right)
	\sum_{s_A+{s} = s} 
	\left( 	
	\begin{array}{ccc}
	\lambda&\ell_A&1\\
	-s&s_A&{s}
	\end{array}\right) \zetalamb_{s,s_A,{s}}(r),
\ee
and the constant $C(\ell_A,{\ell},\lambda)$ is defined by
\st\be
C(\ell_A, {\ell},\lambda) = (-1)^{\ell_A+\lambda+ {\ell}} 
\sqrt{ \frac
{(2\ell_A+1)(2{\ell}+1)(2\lambda+1)}{4\pi}}.
\ee
Since the spin-spherical harmonics obey the orthogonality relation
\st\be
\int_{4\pi} \;_s\!Y^*_{\ell'm'} \;_s\!Y_{\ell m} d \Omega
	= \delta_{\ell\ell'}\delta_{mm'},
\label{ortho}
\ee 
the first magnetic coupling coefficient is just the summation
\st\be
\kappa^{(1)}_{AD}(r) = \int  \delta \B^*_A \cdot \delta \B_D d \Omega
	= \sum_{\lambda,\mu} \left( 
	\delta B_{A,0}^{*\lambda,\mu} \delta B_{D,0}^{\lambda,\mu} 
	+ \delta B_{A,+}^{*\lambda,\mu} \delta B_{D,+}^{\lambda,\mu}
	+ \delta B_{A,-}^{*\lambda,\mu} \delta B_{D,-}^{\lambda,\mu} \right).
\label{kappa1}
\ee

Similar expressions for the coupling coefficients 
$\kappa_{AD}^{(2)}(r)$ and $\kappa_{AD}^{(3)}(r)$
can also be written. In order to calculate $\kappa^{(2)}_{AD}(r)$, the vector
$\xxi \times (\nab \times \B)$ needs to be decomposed into a 
sum over the spin-weighted
spherical harmonics. The decomposition is of the form
\st\be
\cchi_A = \xxi_A \times (\nab \times \B) = 
 \sum_{\lambda,\mu} \left(
\chi_{A,0}^{\lambda,\mu}(r)  \;_0\!Y^*_{\lambda \mu} \unit_0
\;+\; \chi_{A,+}^{\lambda,\mu}(r)  \;_-\!Y^*_{\lambda \mu} \unit_-
\;+\; \chi_{A,-}^{\lambda,\mu}(r)  \;_+\!Y^*_{\lambda \mu} \unit_+\right),
\ee
where the coefficients are
\st\be
\chi_{A,s}^{\lambda,\mu}(r) = \sum_{\ell_A} \sum_{{\ell},{m}}
	C(\ell_A,{\ell},\lambda)  
	\left( 
	\begin{array}{ccc}
	\lambda&\ell_A&{\ell}\\
	\mu&m_A&{m}
	\end{array}\right)
	\sum_{s_A+{s} = s} 
	\left( 	
	\begin{array}{ccc}
	\lambda&\ell_A&{\ell}\\
	-s&s_A&{s}
	\end{array}\right) \chialamb_{s,s_A,{s}}(r).
\ee
The only non-zero functions $\chialamb(r)$ are given by the expressions
\st
\begin{mathletters}
\bea
\chialamb_{0,-1,1}(r) =\frac{1}{r} \fmA (\bzlamb+ (r{\bplamb})_{,r} ) \;,&& \quad
\chialamb_{0,1,-1}(r) = -\frac{1}{r}\fpA (\bzlamb - (r{\bmlamb})_{,r}) \\
\chialamb_{1,0,1}(r) = -\frac{1}{r}\fzA (\bzlamb + (r{\bplamb})_{,r}) \;,&& \quad
\chialamb_{1,1,0}(r) = -\frac{1}{r}\fpA (\bplamb + \bmlamb)\\
\chialamb_{-1,0,-1}(r) = \frac{1}{r}\fzA (\bzlamb - (r{\bmlamb})_{,r} ) \;,&& \quad
\chialamb_{-1,-1,0}(r) = \frac{1}{r}\fmA (\bplamb + \bmlamb).
\eea
\end{mathletters}
The second coupling coefficient $\kappa^{(2)}_{AD}$ is now given by
\st\be
\kappa^{(2)}_{AD}(r) = \int  \cchi^*_A \cdot \delta \B_D d \Omega
	= \sum_{\lambda,\mu} \left( 
	\chi_{A,0}^{*\lambda,\mu} \delta B_{D,0}^{\lambda,\mu} 
	+ \chi_{A,+}^{*\lambda,\mu} \delta B_{D,+}^{\lambda,\mu}
	+ \chi_{A,-}^{*\lambda,\mu} \delta B_{D,-}^{\lambda,\mu} \right).
\ee

Similarly, we define the vector
\st\be
\ggamma_A = \xxi_A \times \B = 
 \sum_{\lambda,\mu} \left(
\gamma_{A,0}^{\lambda,\mu}(r)  \;_0\!Y^*_{\lambda \mu} \unit_0
\;+\; \gamma_{A,+}^{\lambda,\mu}(r)  \;_-\!Y^*_{\lambda \mu} \unit_-
\;+\; \gamma_{A,-}^{\lambda,\mu}(r)  \;_+\!Y^*_{\lambda \mu} \unit_+\right),
\ee
where the expansion coefficients are
\st\be
\gamma_{A,s}^{\lambda,\mu}(r) = \sum_{\ell_A} \sum_{{\ell},{m}}
	C(\ell_A,{\ell},\lambda)  
	\left( 
	\begin{array}{ccc}
	\lambda&\ell_A&{\ell}\\
	\mu&m_A&{m}
	\end{array}\right)
	\sum_{s_A+{s} = s} 
	\left( 	
	\begin{array}{ccc}
	\lambda&\ell_A&{\ell}\\
	-s&s_A&{s}
	\end{array}\right) \gammaalamb_{s,s_A,{s}}(r).
\ee
The non-zero coefficients $\gammaalamb(r)$ have the values
\st
\begin{mathletters}
\bea
\gammaalamb_{0,-1,1}(r) = i \fmA \bplamb \;,&& \quad
\gammaalamb_{0,1,-1}(r) = -i \fpA \bmlamb \\
\gammaalamb_{1,1,0}(r) = i \fpA \bzlamb \;,&& \quad
\gammaalamb_{1,0,1}(r) = -i \fzA \bplamb \\
\gammaalamb_{-1,-1,0}(r) = -i \fmA \bzlamb \;,&& \quad
\gammaalamb_{-1,0,-1}(r) = -i \fzA \bmlamb. 
\eea
\end{mathletters}

The third coupling coefficient is
\st\be
\kappa^{(3)}_{AD}(r) = \int  \ggamma^*_A \cdot \ggamma_D d \Omega
	= \sum_{\lambda,\mu} \left( 
	\gamma_{A,0}^{*\lambda,\mu} \gamma_{D,0}^{\lambda,\mu} 
	+ \gamma_{A,+}^{*\lambda,\mu} \gamma_{D,+}^{\lambda,\mu}
	+ \gamma_{A,-}^{*\lambda,\mu} \gamma_{D,-}^{\lambda,\mu} \right).
\ee

\begin{deluxetable}{llrrrrrr}
\tablecaption{Typical ratios of magnetic field energy to gravitational
and rotational energies for different types of stars, assuming uniform
density and magnetic field. The energy in the magnetic field is
${\cal M} = B_{in}^2 R^3/6$, the gravitational potential energy is $|W| = 3GM^2/5R$
and the rotational energy is $T = MR^2(2\pi/P)^2/5.$
White dwarf data is from \citet{WF00}, Ro AP star data is from \citet{Kur90},
and the data for SGR 0526-66 is from \citet{Dun98}.
\label{t:energy}}
\tablehead{
\colhead{Name}&\colhead{Type}&\colhead{B}&\colhead{Period}
& \colhead{Mass}&\colhead{Radius}
&\colhead{${\cal M}/|W|$}&\colhead{${\cal M}/T$}\\
\colhead{}&\colhead{}&\colhead{(G)}&\colhead{(s)}
& \colhead{($M_\odot$)}&\colhead{(cm)}
&\colhead{}&\colhead{}
}
\startdata
Crab & NS & 1.0e+12 & 3.3e-02 & 1.4 & 1.0e+06 & 5.4e-13 & 8.3e-09 \\
AP stars & RoAp & 1.0e+03 & 1.0e+06 & 2.0 & 1.4e+11 & 1.0e-10 & 7.4e-07 \\
AM Her & WD & 1.3e+07 & 1.1e+04 & 0.7 & 1.0e+09 & 3.6e-10 & 3.2e-04 \\
PG 1015+14 & WD & 1.6e+08 & 5.9e+03 & 0.7 & 1.0e+09 & 5.5e-08 & 1.4e-02 \\
PG 1031+234 & WD & 1.0e+09 & 1.2e+04 & 0.7 & 1.0e+09 & 2.1e-06 & 2.3e+00 \\
SGR 0526-66 & NS & 1.0e+15 & 8.0e+00 & 1.4 & 1.0e+06 & 5.4e-07 & 4.9e+02 \\     \enddata
\end{deluxetable}

\begin{deluxetable}{lll}
\tablecaption{The numerical values of $\kappa_{AA}$ for force-free magnetic 
field. The data are calculated for $\mu=3.51835$ and $\frac{B_P}{B_{in}}=0.1$.
The coupling coefficients are written in the form
$\kappa_{AA} = a_{AA} + b_{AA} \sin^2\alpha$ where $\alpha$ is the angle
between the star's spin axis and the magnetic field's symmetry axis.
\label{t:kappa}}
\tablehead{
\colhead{$\ell_A$}&\colhead{$a_{AA}$}&\colhead{$b_{AA}$}
}
\startdata
1&-4.2182e-01 & 2.1091e-01 \\
2&-5.9812e-01 & 2.5302e-02 \\
3&-1.5734e+00 & 7.8447e-03 \\
4&-4.1159e+00 & 2.9866e-01 \\
5&-9.2792e+00 & 1.0233e+00 \\
6&-1.8367e+01 & 2.2962e+00 \\
7&-3.2933e+01 & 4.2251e+00 \\
8&-5.4780e+01 & 6.9128e+00 \\
9&-8.5963e+01 & 1.0458e+01 \\
10&-1.288e+02 & 1.4955e+01 
\enddata
\end{deluxetable}

\begin{figure}
\figurenum{1}
\label{fig1}
\plotone{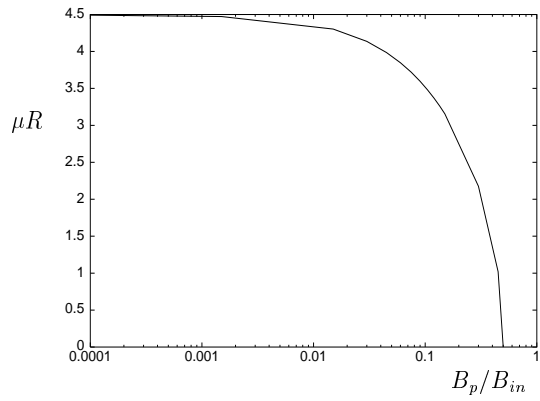}
\caption{Plot of $\mu R$ versus $B_p/B_{in}$, for the force-free
magnetic field. In this plot
$R$ is the radius of the star, $\mu$ is the proportionality
constant appearing in equation (\ref{ff}) and $B_p/B_{in}$
is the ratio of external to internal magnetic fields. 
This
plot was made by numerically finding the solution of equation (\ref{j1eq}).}
\end{figure}

\begin{figure}
\figurenum{2}
\label{fig2}
\plotone{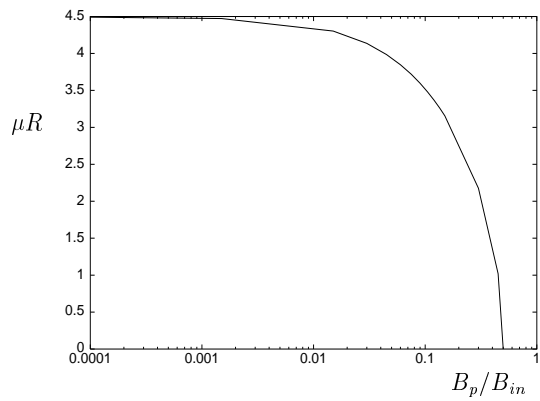}
\caption{The magnetic frequency correction 
for an $\ell_A=m_A=2$ r-mode  as a function of $B_p/B_{in}$, for the force-free
magnetic field. The frequency correction is written 
in the form $\kappa_{AA} = a_{AA} + b_{AA} \sin^2\alpha$, and the 
constants $a_{22}$ and $b_{22}$ are plotted. The solid curve is the
graph of the function $a_{22}$ and the dot-dashed line is the 
graph of $b_{22}$.
}
\end{figure}


\begin{thebibliography}  

\bibitem[Andersson(1998)]{And98}
	Andersson, N. 1998, \apj, 502, 708

\bibitem[Bigot et al.(2000)]{Big00}
	Bigot, L., Provost, J., Berthomieu, G., 
	Dziembowski, W. A., \& Goode, P. R. 2000,
	Astron. Astrophys., 356, 218

\bibitem[Boriakoff(1976)]{Bor76}
	Boriakoff, V. 1976, \apj, 208, L43

\bibitem[Campbell(1971)]{Cam71} Campbell, W. B. 1971,
	J. Math. Phys., 12, 1763

\bibitem[Carroll et al.(1986)]{Car86}
	Carroll, B. W., Zweibel, E. G., Hansen, C. J.,
	McDermott, P. N., Savedoff, M. P., Thomas, J. H.,
	\& Van Horn, H. M. 1986, \apj, 305, 767

\bibitem[Chandrasekhar \& Fermi(1953)]{CF53}
	Chandrasekhar, S. \& Fermi, E. 1953,
	\apj, 118, 116

\bibitem[Cordes et al(1990)]{Cor90}
	Cordes, J. M., Weisberg, J. M., Hankins, T. H. 1990,
	\aj, 100, 1882

\bibitem[Cox(1980)]{Cox80}
	Cox, J. P. 1980, 
	{\em Theory of Stellar Pulsation}, 
	(Princeton, NJ: Princeton University Press)

\bibitem[Duncan(1998)]{Dun98} Duncan, R. C. 1998,
	\apj, 498, L45

\bibitem[Dyson \& Schutz(1979)]{DS79}
	Dyson, J. \& Schutz, B. F. 1979,
	Proc. R. Soc. Lond. A 368, 389

\bibitem[Dziembowski \& Goode(1996)]{DG96}
	Dziembowski, W.A., \& Goode, P. R. 1996,
	\apj, 458, 338

\bibitem[Edmonds(1974)]{Edm74} Edmonds, A. R. 1974,
	{\em Angular Momentum in Quantum Mechanics},
	2nd Edition with corrections,
	(Princeton, NJ: Princeton University Press)

\bibitem[Ferraro \& Plumpton(1966)]{FP66} 
        Ferraro, V. C. A., \& Plumpton, C. 1966,
        {\it An Introduction to Fluid Mechanics}, 
	2nd Edition, (Oxford, UK: Oxford University
        Press) pp. 37-46 and 62-65

\bibitem[Friedman \& Morsink(1998)]{FM98}
	Friedman, J. L. \& Morsink, S. M. 1998, \apj, 502, 714

\bibitem[Friedman \& Schutz(1978a)]{FS78} Friedman, J. L. \&
	Schutz, B. F. 1978a, \apj, 221, 937 

\bibitem[Friedman \& Schutz(1978b)]{FS78b} Friedman, J. L. \&
	Schutz, B. F. 1978b, \apj, 222, 281 

\bibitem[Gautschy \& Saio(1996)]{GS96}
	Gautschy, A., \& Saio, H. 1996,	
	\araa, 34, 551

\bibitem[Ho \& Lai(2000)]{HL00}
	Ho, W. C. G. \& Lai, D. 2000, \apj, 543, 386

\bibitem[Jackson(1975)]{Jac75} Jackson, J. D. 1975,
      {\it Classical Electrodynamics}, 2nd Edition, 
	(New York, NY: John Wiley \& Sons)

\bibitem[Kouveliotou(1998)]{Kou1998}
         Kouveliotou, C. et al. 1998, Nature, 393, 235

\bibitem[Kurtz(1990)]{Kur90} Kurtz, D. W. 1990, \araa, 28, 607

\bibitem[Ledoux \& Simon(1957)]{LS57}
	Ledoux, P. \& Simon, R. 1957, Ann. Astrophys. 20, 185
	
\bibitem[Lindblom et al.(1998)]{LOM98}
	Lindblom, L., Owen, B. J., \& Morsink, S. M. 1998, 
	\prl, 80, 4843

\bibitem[Lockitch \& Friedman(1999)]{LF99}
	Lockitch, K. H., \& Friedman, J. L. 1999, \apj, 521, 764

\bibitem[Malkus(1967)]{Mal67}
	Malkus, W. V. R. 1967, J. Fluid Mech., 28, 793

\bibitem[Rezzolla et al.(2000)]{Rez00}
	Rezzolla, L., Lamb, F. K., \& Shapiro, S. L. 2000,
	\apj, 531, L141

\bibitem[Rezzolla et al.(2001a)]{Rez01a}
	Rezzolla, L., Lamb, F. K., Markovi\'{c}, D. \& Shapiro, S. L. 2001a,
	\prd, 64, 104013

\bibitem[Rezzolla et al.(2001b)]{Rez01b}
	Rezzolla, L., Lamb, F. K., Markovi\'{c}, D. \& Shapiro, S. L. 2001b,
	\prd, 64, 104014

\bibitem[Rieutord et al.(2001)]{RGV01}
	Rieutord, M., Georgeot, B., \& Valdettaro, L. 2001,
	J. Fluid Mech., 435, 103

\bibitem[Schenk et al.(2002)]{Sch01} Schenk, A. K., Arras, P., 
	Flanagan, \'{E}. \'{E}.,
        Teukolsky, S. A., \& Wasserman, I., 2002, 
	\prd, 65, 024001

\bibitem[Schmidt \& Grauer(1997)]{SG97} Schmidt, G. D., \& Grauer, A. D. 1997,
	\apj, 488, 827

\bibitem[Shibahashi \& Takata(1993)]{ST93}
	Shibahashi, H. \& Takata, M. 1993, \pasj, 45, 617

\bibitem[Takata \& Shibahashi(1994)]{TS94}
	Takata, M. \& Shibahashi, H. 1994, \pasj, 46, 301

\bibitem[Takata \& Shibahashi(1995)]{TS95}
	Takata, M. \& Shibahashi, H. 1995, \pasj, 47, 219


\bibitem[Thompson \& Duncan(1995)]{TD95} Thompson, C., \& Duncan, R. C. 1995,
	\mnras, 275, 255

\bibitem[Unno et al.(1989)]{Unn89}
	Unno, W., Osaki, Y., Ando, H., Saio, H., \& 
	Shibahashi, H. 1989, 
	{\em Nonradial Oscillations of Stars}, 
	(Tokyo, Japan: University of Tokyo Press) 

\bibitem[Van Horn(1980)]{VH80}
	Van Horn, H. M. 1980, \apj, 236, 899

\bibitem[Wickramasinghe \& Ferrario(2000)]{WF00} 
	Wickramasinghe, D. T., \& Ferrario, L. 2000,
	\pasj 112, 873
 

\end{thebibliography}
\end{document}